\documentclass{article}

\usepackage[final]{neurips_2020}

\usepackage[]{natbib}
\setcitestyle{authoryear,round,citesep={;},aysep={,},yysep={;}}

\usepackage[utf8]{inputenc} 
\usepackage{hyperref}       
\usepackage{url}            
\usepackage{booktabs}       
\usepackage{amsfonts}       
\usepackage{amsmath}        
\usepackage{nicefrac}       
\usepackage{microtype}      
\usepackage{xcolor}

\usepackage{graphicx}       

\title{Evolution Is All You Need: Phylogenetic Augmentation for Contrastive Learning}

\author{%
  Amy X. Lu \\
  University of Toronto \\
  Vector Institute\\
  \texttt{amyxlu@cs.toronto.edu} \\
  \And
  Alex X. Lu \\
  University of Toronto \\
  \texttt{alexlu@cs.toronto.edu} \\
  
  \And
  Alan Moses \\
  University of Toronto \\
  \texttt{alan.moses@utoronto.ca} \\

}

\begin{document}

\maketitle

\begin{abstract}
Self-supervised representation learning of biological sequence embeddings alleviates computational resource constraints on downstream tasks while circumventing expensive experimental label acquisition. However, existing methods mostly borrow directly from large language models designed for NLP, rather than with bioinformatics philosophies in mind. Recently, contrastive mutual information maximization methods have achieved state-of-the-art representations for ImageNet. In this perspective piece, we discuss how viewing evolution as natural sequence augmentation and maximizing information across phylogenetic ``noisy channels" is a biologically and theoretically desirable objective for pretraining encoders. We first provide a review of current contrastive learning literature, then provide an illustrative example where we show that contrastive learning using evolutionary augmentation can be used as a representation learning objective which maximizes the mutual information between biological sequences and their conserved function, and finally outline rationale for this approach.

\end{abstract}

\section{Introduction}

Self-supervised learning representation learning of biological sequences aims to capture meaningful properties for downstream analyses, while pretraining only on labels derived from the data itself. Embeddings alleviate computational constraints, and yield new biological insights from analyses in a rich latent space; to do so in a self-supervised manner further circumvents the expensive and time-consuming need to gather experimental labels. Though recent works have successfully demonstrated the ability to capture properties such as fluorescence, pairwise contact, phylogenetics, structure, and subcellular localization, these works mostly use methods designed for natural language processing (NLP) \citep{yang2018learned,bepler2019learning,riesselman2019accelerating,rives2019biological,alley2019unified,heinzinger2019modeling,elnaggar2019end,gligorijevic2019structure,armenteros2020language,madani2020progen,elnaggar2020prottrans}. This leaves open the question of how best to design self-supervised methods which align with biological principles.

Recently, contrastive methods for learning representations achieve state-of-the-art results on ImageNet~\citep{oord2018representation,henaff2019data,tian2019contrastive,he2019momentum,chen2020simple}. Two ``views" $v_1$ and $v_2$ of an input are defined (e.g. two image augmentation strategies), and the contrastive objective is to distinguish one pair of ``correctly paired" views from $N-1$ ``incorrectly paired" dissimilar views. This incentivizes the encoder to learn meaningful properties of the input, while disregarding nuisance factors. Theoretically, it can be shown that such an objective maximizes the lower-bound on the mutual information, $I(v_1, v_2)$ \citep{poole2019variational}.

In this piece, we first provide a review of current contrastive learning literature for obtaining representations in non-biological modalities. Then, we propose that \textit{molecular evolution} is a good choice of augmentation to provide ``views” for contrastive learning in computational biology, from both the theoretical and biological perspectives. Finally, we illustrate how evolutionary augmentation can be used to optimize a deep neural network encoder to preserve the information in biological sequences that pertains to their function.

\section{Contrastive Learning for Mutual Information Maximization}

To encourage the development of novel contrastive methods for biological applications, we provide a broad overview of existing contrastive methods. Section \ref{sec:simclr_bio} describes one such method, SimCLR, in greater detail.

\label{sec:background}
\subsection{Contrastive Learning and Mutual Information Estimation}
The InfoMax optimization principle \citep{linsker1988self} aims to find a mapping $g$ such that the Shannon mutual information between the input and output is maximized, i.e. $\max_{g \in \mathcal{G}} I(X; g(X))$. Recent works revive this principle as a representation learning objective to train deep encoders as $g$, and yield empirically desirable representations in the modalities of imaging \citep{oord2018representation,hjelm2018learning,bachman2019learning,tian2019contrastive,henaff2019data,lowe2019putting,he2019momentum,chen2020simple,tian2020makes,wang2020understanding}, text \citep{riviere2020unsupervised,oord2018representation,kong2019mutual}, and audio \citep{lowe2019putting,oord2018representation}.

Most follow a variation of this optimization objective: given input $x$, and transformations $t_1$ and $t_2$, define $v_1 = t_1(x)$ and $v_2 = t_2(x)$ as two different ``views" of $x$. These ``transformations" can be parameterless augmentations \citep{chen2020simple}, or another neural network summarizing global information \citep{hjelm2018learning,oord2018representation}. Further, define encoder(s) $g_1$ and $g_2$, and latent representations $z_1 = g_1(v_1)$ and $z_2 = g_2(v_1)$. The encoder mappings may be constrained by $\mathcal{G}_1$ and $\mathcal{G}_2$ (e.g. architecturally). In some works, $g_1$ and $g_2$ may share some \citep{hjelm2018learning} or all \citep{chen2020simple} parameters. The goal is to find encoder mappings which maximize the (estimated) mutual information between the outputs: 

\begin{equation}
    \label{eq:infomax_rep}
    \max_{g_1 \in \mathcal{G}_1, g_2 \in \mathcal{G}_2} I' (g_1(v_1); g_2(v_2)) 
\end{equation}

This objective is shown to lower-bound the true InfoMax objective \citep{tschannen2019mutual}. Perhaps the most widely adapted estimator is the InfoNCE estimator, which provides an unnormalized lower bound on the mutual information by optimizing the objective \citep{oord2018representation}:

\begin{equation}
    \label{eq:infonce}
    \mathcal{L}_{NCE} := -\mathbb{E}_{v_1,v^-_2,v^+_2} \Big[ \log \frac{\exp(f(g_1(v_1), g_2(v^+_2)))}{\exp(f(g_1(v_1),g_2(v_2^+))) + \sum_{j=1}^{N-1} \exp(f(g_1(v_1),g_2(v^-_{2_{j}})))} \Big],
\end{equation}

where $(v_1, v_2^+) \sim p(v_1, v_2)$ is a ``real" pair of views drawn from their empirical joint distribution, and we draw negative samples $v_2^- \sim p(v_2)$ from the marginal distribution to form $N-1$ ``fake" pairs. $N$ denotes the total number of pairs (and, in practice, often refers to the batch size). In \citet{arora2019theoretical}, losses in this general form are termed ``contrastive learning".

We see that Equation \ref{eq:infonce} is a cross-entropy which distinguishes one positive pair from $N-1$ negative pairs, where $f$ is a ``critic" classifier (reminiscent of adversarial learning), and should learn to return high values for the ``real" pair. As is common in deep learning, the expectation is calculated over multiple batches. 
For a more detailed discussion of the connection between the InfoNCE loss, the InfoMax objective for representation learning, and other mutual information estimators, see Appendix \ref{sec:app_mi}.

\subsection{Choice of ``Views" in Contrastive Learning Literature}
Existing works select ``views" of the input in different ways. These include using different time steps of an audio or video sequence \citep{oord2018representation,sermanet2018time} or using different patches of the same image \citep{oord2018representation,henaff2019data,hjelm2018learning,bachman2019learning}. Recently, contrastive learning between local and sequentially-global embeddings is used to establish representations for proteins \citep{lu2020self}. Augmentations are an oft-used strategy for constructing different views \citep{hu2017learning,he2019momentum,chen2020simple}, sometimes applied in conjunction with image patching \citep{henaff2019data,bachman2019learning}.

In this work, we argue that using evolution as a sequence augmentation strategy is a biologically and theoretically desirable choice to construct views. Previous work have explored evolutionary conservation as a means of sequence augmentation during training, such as augmenting a HMM using simulated evolution \citep{kumar2009augmented}, or generating from a PSSM \citep{asgari2019deepprime2sec}. Other methods include using generative adversarial networks (GANs) for -omics data augmentation \citep{eftekhar2020prediction,marouf2020realistic} or injecting noise by replacing amino acids from an uniform distribution \citep{koide2018neural}. For genomic sequences, augmentations can be formed using reverse complements and extending (or cropping) genome flanks \citep{cao2019simple,kopp2020deep}.

\section{Evolution as Sequence Augmentation for Contrastive Learning}
\label{sec:simclr_bio}

As a motivating example of how phylogenetic augmentation can be used in contrastive learning, we specifically describe how SimCLR \citep{chen2020simple} can be applied to biological sequences, though specific details (e.g. choice of critic, encoding process, etc.) can be adapted following details in Section \ref{sec:background}.

\begin{figure}[h!]
    \centering
    \includegraphics[scale=0.3]{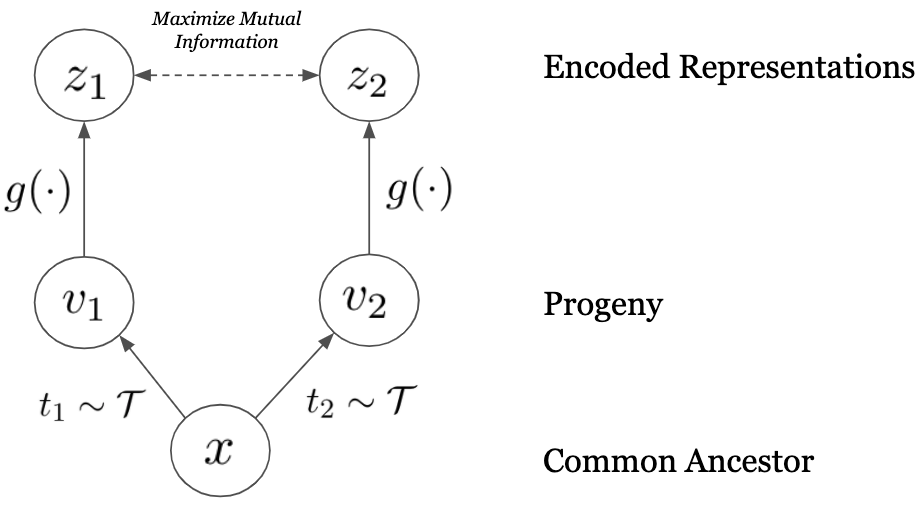}
    \caption{SimCLR \citep{chen2020simple} can be re-casted as a phylogenetic tree where augmentations are evolution. In the original \citet{chen2020simple} paper, $x$ is an input image, and two image augmentation methods, $t_1$ and $t_2$, are sampled from a set of image augmentation methods $\mathcal{T}$, to produce image augmentation $v_1$ and $v_2$, which are then passed into a trainable encoder $g(\cdot)$ (i.e. $g_1$ and $g_2$ share parameters entirely). In conceptualizing evolution as an augmentation strategy, $x$ can be viewed as a common ancestor, while $\mathcal{T}$ denote possible evolutionary trajectories, characterized by different evolutionary distance, mutation and genetic drift, and $t_1, t_2$ are two example trajectories that lead to  $v_1$ and $v_2$, sampled from a set of homologs. Note that notations are adapted from the original SimCLR paper for consistency with the current work.}
    \label{fig:simclr_phylogenetic}
\end{figure}

As outlined in Figure \ref{fig:simclr_phylogenetic}, homologous sequences can be considered as ``evolutionary augmented views" of a common ancestor, $x$. Sequences $v_1$ and $v_2$ are encoded by an encoder $g(\cdot)$ to obtain embeddings $z_1$ and $z_2$. The pair of embeddings augmented from the same ancestor -- that is, embeddings of homologous sequences -- will be the positive pair $(v_1, v^+_2) \sim p(v_1, v_2)$. To sample negative samples $\{v^-_{2_{j}}\}^{N-1}_{j=1}$ from $p(v_2)$, we can draw negatives from all non-homologous sequences.

The key idea is that properties of the ancestral sequence that were important for its biological function will be preserved in both descendants (i.e. views). By training the encoder to project these to nearby locations in the latent space, we ensure that proximity in the latent space corresponds to similar biological functions without explicit labels during pretraining, analogous to how SimCLR learns semantic content without image labels. Practically, existing bioinformatics databases of homologs can be used; since contrastive learning uses pairs of positive examples, the number of training examples for pretraining increases quadratically with the number of homologs. We see that contrastive learning frameworks such as SimCLR can be directly adapted to capture phylogenetic principles.

\section{Why Evolution as Biological Sequence Augmentation?}
\subsection{Invariant Representations Across Evolutionary ``Noisy-Channels" Mirrors Comparative Genomics}
\textit{Biological sequences are vehicles for information transmission}. As such, information theoretic principles are directly applicable to biological sequence analyses, and therefore, this may be more a more powerful approach than  methods based on the analogy with  natural language \citep{alley2019unified,rives2019biological,rao2019evaluating,elnaggar2020prottrans}.

The analogy between molecular evolution and noisy-channel coding is well-rooted in prior work \citep{gatlin1972information,mackay2003information,vinga2014information,kuruoglu2017information}: DNA dictates information transmission across generations, which must be transferred through a noisy “mutation and drift channel". Further, as noted in \citet{kimura1961natural}, as the genotype-to-phenotype manifestation is information transfer, and genomic information is passed down by heredity, we may view functional phenotypes as ``decoded" information that was transmitted from a common ancestor via molecular evolution. Drawing from these writings, we argue that using maximizing mutual information across homologs is a good proxy for structure and function \citep{adami2000evolution}, which are the central aims for biological sequence embeddings \citep{rao2019evaluating}.


Even without relying on the mutual information estimation interpretation of the InfoNCE loss, the contrastive learning objective directly encourage representational invariance to shared features across views \citep{chen2020simple}. Therefore, in using phylogenetic relationships to create views, learned representations directly capture the philosophy of \textit{evolutionary conservation} in comparative genomics: functional elements will be preserved in comparisons of related sequences, while non-functional sequences will decay. Hence, functional elements in biological sequences can be identified through sequence comparisons \citep{hardison2003comparative}. This is perhaps the most successfully employed presumption in bioinformatics \citep{eddy1998profile,altschul1990basic}. 

We therefore argue that InfoMax-based deep learning on evolutionary augmentation has two attractive features from the biological perspective: (1) Molecular evolution and the genotype-to-phenotype relationship has a clear analogy to information transmission; and (2) contrastive learning in this setting encourages agreement between important features across evolutionary views (homologous sequences), which directly mirrors comparative genomics.

\subsection{Evolutionary Augmentation is a Theoretically Desirable View}
\label{sec:mi_in_bio}

\citet{tian2020makes} proposes the ``InfoMin" principle for selecting optimal views. The authors theoretically and empirically demonstrate that good views should have \textit{minimize} their shared MI while keeping \textit{task-relevant} information intact for downstream uses. More formally, for a downstream classification task $C$ to predict label $y \in \mathcal{Y}$ from $x$, the optimal representation $z^* = g_1(x)$ is the is the minimal sufficient statistic for task $C$, such the representation is as useful as access to $x$ while disregarding all nuisance in $x$ \citep{tian2020makes,soatto2014visual}. Then, the optimal views of task $C$ is $(v_1^*, v_2^*) = \min_{v_1;v_2} I(v_1;v_2)$, subject to $I(v_1;y) = I(v_2;y) = I(x;y)$. With optimal views $(v^*_1, v^*_2)$, the subsequently learned representations $(z^*_1, z^*_2)$ are optimal for task $C$.\footnote{The optimal property of representations $(z^*_1, z^*_2)$ assumes access to a ``minimally sufficient encoder" which serve as a minimal sufficient statistic of the input \citep{soatto2014visual}. More formally, a ``sufficient encoder" $g_{\text{sufficient}}$ require that $g_{\text{sufficient}}(v_1)$ has kept all information about $v_2$ in $v_1$, and a ``minimal sufficient encoder" $g_1 \in \mathcal{G}_{\text{sufficient}}$ discards all irrelevant ``nuisance" information such that $I(g_1(v_1); v_1) \leq I(g_{\text{sufficient}}(v_1);v_1), \forall g_{\text{sufficient}} \in \mathcal{G}_{\text{sufficient}}$.}

There are two implications in adapting the InfoMin principle to biology which render evolutionary augmentations desirable. Firstly, sampling evolutionary trajectories $t_1, t_2 \sim \mathcal{T}$ to create $v_1=t_1(x)$ and $v_2=t_2(x)$ provide a simple way to reduce $I(v_1, v_2)$ by selecting paired views $(v_1, v_2^+)$ with a greater phylogenetic distance between them. Secondly, note that in order to choose views based on the InfoMin principle, access to labels $y \in \mathcal{Y}$ and knowledge of task $C$ is needed. In fact, \textit{supervised} contrastive learning \citep{khosla2020supervised} empirically yields improved results by explicitly sampling negatives from a different downstream class. If given labels for a downstream biological task of interest (e.g. remote homology), one can explicitly negative sample from dissimilar classes (e.g. different folds); however, owing to the difficult label-acquisition process and open-ended nature of biological questions, access to $\mathcal{Y}$, or even task $C$, may not be always possible. Further, it is often desirable for biological sequence embeddings to be ``universal representations" \citep{alley2019unified} and applicable for a variety of downstream tasks \citep{rao2019evaluating}. As noted in Section \ref{sec:mi_in_bio}, evolutionary conservation is a good proxy for many tasks of interest (e.g. structure and function).


Thus, we see that in transferring theoretical results for optimal view selection to the biological setting, evolution as augmentation is desirable, as: (1) It is easy to control shared mutual information between views; and (2) evolutionary conservation is a good semantic proxy for downstream labels, and implicitly performs \textit{supervised} contrastive learning while still circumventing expensive experimental label gathering. Hence, it may be best considered a general strategy for weakly-supervised contrastive learning.

\section{Conclusion}
Current methods for self-supervised representation learning in biology are mostly adapted from NLP methods. Contrastive learning achieves state-of-the-art results in the image modality, and has a desirable theoretical property of being a lower-bound estimator of mutual information. We demonstrate how evolution can be used as a sequence augmentation strategy for contrastive learning, and provide justifications for doing so from biological and theoretical perspectives. More generally, data augmentation is a critical preprocessing step in many image analysis applications of deep learning, but is it less clear how to augment data for biological sequence analysis. As research in applications of deep learning in biology expand, we hope the view of evolution as augmentation will guide the ideation of deep learning methods in computational biology. 

\bibliography{main}

\appendix
\section{InfoMax Principle and Mutual Information Estimation for Representation Learning}
\label{sec:app_mi}

\subsection{Applying InfoMax to Representation Learning}
\label{sec:app_infomax_for_rep}

Using InfoMax for representation learning extends as far back as ICA \citep{bell1995information}. As described in Equation \ref{eq:infomax_rep}, in recent years, works typically maximize mutual information of two \textit{encoded} ``views" of an input (e.g. different patches of an image, or augmentations). By the data processing inequality, \citet{tschannen2019mutual} show that:

\begin{equation}
    \label{eq:infomax_bound}
    I(g_1(v_1); g_2(v_2)) \leq I(x; g_1(v_1), g_2(v_2)),
\end{equation}

such that maximizing Equation \ref{eq:infomax_rep} is equivalent to maximizing a lower bound on the true InfoMax objective. This ability to maximize the mutual information in the latent embedding space rather than directly between the input and the encoded output (as per the original InfoMax formulation) has a few advantages \citep{tschannen2019mutual}: MI is difficult to estimate in high dimensions, and this formulation enables MI estimation in a lower-dimensional space; furthermore, the ability to use creative encoders for $\mathcal{G}$ can accommodate specific modelling needs and data intricacies.
data intricacies. 

\subsection{InfoNCE Estimator}
\label{sec:app_infonce}
InfoNCE is one of many mutual information estimators, and following the rationale in Section \ref{sec:app_infomax_for_rep}, the original \citet{oord2018representation} paper does this estimation in the embedding space. For the InfoNCE loss (Equation \ref{eq:infonce}) which estimates $I(z_1;z_2) = I(g_1(v_1);g_2(v_2))$ in Equation \ref{eq:infomax_bound}, the optimal critic function is $f^*(z_1, z_2) = \frac{p(z_2|z_1)}{p(z_1)}$ \citep{oord2018representation}. Inserting this in the InfoNCE loss function (Equation \ref{eq:infomax_rep}) and rearranging, we have the bound \citep{oord2018representation,poole2019variational}:

\begin{equation}
    \label{eq:infonce_bound}
    I(z_1, z_2) \geq \log(N) - \mathcal{L^{*}_{\text{NCE}}}
\end{equation}

where $N$ is the number of samples. From Equation \ref{eq:infonce_bound}, note that the bound is tight when: (1) We use more samples for $N$ which increases the $\log(N)$ term; and (2) we have a better $f$ which results in a lower $\mathcal{L_{\text{NCE}}}$. Empirically, most works corroborate the former theoretical observation regarding $N$ (exceptions being \citet{arora2019theoretical,lu2020self}), while the latter observation regarding $f$ does not usually hold, as will be further discussed in Section \ref{sec:app_mutual_information_estimators}.

The contrastive nature of the InfoNCE loss stems from its direct adaptation of the noise-contrastive estimation (NCE) method \citep{gutmann2010noise}. Noise-contrastive estimation was originally proposed for the problem of estimating parameters for unnormalized statistical models in high dimensions, by reducing the problem to simply estimating logistic regression parameters to distinguish between observed data and noise. In InfoNCE, the distinction is made between ``similarity scores", as scored by critic $f(z_1, z_2)$, for one positive pair and $N-1$ negative pairs of encoded views.

\subsection{Other Mutual Information Estimators}
\label{sec:app_mutual_information_estimators}

The InfoNCE estimator is one of many approaches which builds on advancements in variational methods to create differentiable and tractable sample-based mutual information estimators in high dimensions \citep{donsker1983asymptotic,barber2003algorithm,nguyen2010estimating,alemi2016deep,belghazi2018mine,oord2018representation,hjelm2018learning}. Many of these estimations involve a ``critic" classifier, $f$, which can be as simple as a bilinear model \citep{oord2018representation,henaff2019data,tian2019contrastive} or dot-product \citep{chen2020simple,lu2020self}, or a model accepting a concatenation of two views as input. There may be a different $f$ for each view \citep{oord2018representation}, or a global $f$ \citep{hjelm2018learning}. Usually, $f$ is trained jointly with $g_1$ and $g_2$.

The aim of $f$ is often to approximate the unknown densities $p(B)$ and $p(B|A)$, or density ratios $\frac{p(A|B)}{p(B)} = \frac{p(B|A)}{p(A)}$ \citep{poole2019variational}.
Intuitively, if $I(A,B)$ is high, then $f$ should intuitively be able to easily assign high probabilities to those samples drawn from $p(A,B)$ \citep{tschannen2019mutual}. The InfoNCE estimator reduces variance as compared to other estimators, by depending on multiple samples, but trades off bias to do so \citep{poole2019variational}.

Importantly, it should be noted that whether the empirical success of the InfoNCE loss should be attributable to mutual information estimation has been questioned \citep{poole2019variational,tschannen2019mutual}, instead attributing success to geometric properties in the latent space \citep{wang2020understanding}. For example, a higher-capacity $f$ should increase tightness of the bound, as noted in Section \ref{sec:app_infonce}, yet hinders performance \citep{tschannen2019mutual}. The development of MI-estimators useful for neural network training -- and demystifying their empirical success -- remains an active area of research. For the purposes of ideas in this work, we note that SimCLR-like contrastive losses itself intuitively maximizes agreement between views in the representation space without relying on the mutual information framing of the loss \citep{chen2020simple}, and hence the connection to comparative genomics still hold.

\end{document}